\documentstyle[draft]{l-aa}

\thesaurus{}

\newcommand{\etal}{{\it et al.}}
\newcommand{\kms}{\mbox{\ km\ s$^{-1}$}}
\newcommand{\msunyr}{\mbox{M$_{\odot}$\thinspace yr$^{-1}\;$}}
\newcommand{\msun}{\mbox{$M_{\odot}\;$}}
\newcommand{\mstar}{\mbox{$M_{\ast}\;$}}
\newcommand{\rsun}{\mbox{$R_{\odot}\;$}}
\newcommand{\rstar}{\mbox{$R_{\ast}\;$}}
\newcommand{\lsun}{\mbox{$L_{\odot}\;$}}
\newcommand{\lstar}{\mbox{$L_{\ast}\;$}}
\newcommand{\ltappeq}{\mathrel{\hbox{\rlap{\hbox{\lower4pt\hbox{$\sim$}}}\hbox{$<$}}}}

\newcommand{\mnras}{{\sl MNRAS}}
\newcommand{\apj}{{\sl ApJ}}

\newcommand{\araa}[1]{{\sl Ann. Rev. Astr. Astrophys.} {\bf #1}}
\newcommand{\aanda}{{\sl A\&A}}

\newcommand{\pasp}[1]{{\sl PASP} {\bf #1}}

\newcommand{\mdot}{\mbox{$\stackrel{.}{\textstyle M}$}}
\newcommand{\aj}[1]{{\sl Astron. J.} {\bf #1}}

\title{On the warping of Be star discs}
\author{John M. Porter}
\institute{Astrophysics Research Institute,
School of Engineering, Liverpool John Moores University, \\ 
Byrom Street, Liverpool L3 3AF, UK \\
(email : {\tt jmp@astro.livjm.ac.uk}) }

\date{Recieved 13 May 1998 / Accepted 26 May 1998}

\begin{document}
\maketitle

\begin{abstract}
The theory of radiatively-induced warps in accretion discs is applied
to the discs of Be stars. It is found that these discs may
develop warps in their inner regions, although once the warp
amplitude is large enough then the interaction between the disc and
fast radiatively-driven wind will determine its evolution.
The warping is shown to be more important for later than
earlier B stars.
Although the interaction of the fast-wind with the disc will limit the
amplitude of the warp, it cannot drive the warp radially outwards, and
so the radial evolution of the warp depends on the dominant advective
process within the disc.

Typical timescales associated with growing modes are shown to be
short, of the order of days-weeks, although these are not likely to be the
timescales inferred from observations of line-profile variations which
are much longer, of the order of years.

\keywords{stars: emission-line, Be -- stars: rotation -- circumstellar
matter}

\end{abstract}

%-----------------------------------------------------------------------%
\section{Introduction}

Be stars are accepted to have a dense, slowly expanding disc in their
equatorial planes and a fast radiatively-driven wind over their polar
regions (see Slettebak 1988 for a review).
The emission line profiles generated in the disc have been monitored
for many stars and are found to vary (e.g. Dachs 1987). 
Long-term variations in the ratio of the violet to red components of a
line, $V/R$, are attributed to $m = 1$ density waves in a {\em Keplerian}
disc (Okazaki 1991, Papaloizou \etal\ 1992). 
Hanucshik (1996) also concludes that the disc is rotationally
supported.
Using the assumption that the 
disc is rotating at Keplerian velocities around
the central star and is present throughout most of the star's life,
Porter (1998) has recently showed that the radial velocity at the 
inner edge of the disc must be small ($v_r \ltappeq 0.01\kms$). 
The conclusion of these studies is that 
the disc material orbits many times before it travels a significant
radial distance -- similar conditions to the material in an accretion
disc. 

Pringle (1996), Malony, Bagelman \& Pringle (1996) and Armitage \&
Pringle (1997) have recently examined the possibility that a warp may
be induced in an accretion disc due to the incident radiation field
from the central object. They find that if the accretion disc is
optically thick at the wavelength at which the disc emission
is maximal then the disc may become unstable.

Here it is recognised that the physical situation examined by Pringle
and co-workers and that of the Be star disc are similar:
the Be star disc is being illuminated from both sides by the
central B star; the Be star disc material orbits many times before
moving a signifiant radial distance, and observational as well as
theoretical studies have shown that at some radius the Be star discs
are optically thick 
(e.g. Chokshi \& Cohen 1987, Kastner \& Mazzali 1989, Marlborough \etal\ 1997).
Recently, two Be stars have been examined and have shown observational
charactertistics of a warped disc (Hummel, 1998).
In this paper it is attempted to apply instability criteria 
for radiatively-driven warping to the discs
thought to exist around Be stars. 

In \S2 the disc structure is examined and the optically thick regions
are identified.
The criterion for the radiatively-driven warp instability to arise in
the disc is then examined in \S3. The evolution of a warp
interacting with fast radiatively-driven wind is considered in \S4. 
This is discussed in \S5, and conclusions given in \S6.

%-----------------------------------------------------------------------%
\section{Disc structure}

The discs around Be stars have often been modelled with the ``disc
model'' due to Waters (1986). This assumes that the disc density
follows a power law in radius and, from mass conservation within
the disc, that the radial
velocity also has a power law form. 
Typically, the disc has a constant opening angle $\theta$.
Although this model is
emperical, is is very successful in reproducing many of the
observations of Be star discs. (e.g. Waters 1986, Waters, Cot\'{e} \&
Lamers 1987)

However, theoretical models
of disc formation and evolution have been less
successful. Models have been developed based on magnetism (Poe \&
Friend 1986),  
changing wind line-driving parameters (Chen, Marlborough \& Waters 1992), 
wind compression (Bjorkman \& Cassinelli 1993, Porter 1997), 
pulsation (Willson 1986) 
and viscous decretion (Lee, Saio \& Osaki 1991).
None of these models may simultaneously account for many of the
observed features of discs. 
Porter's (1998) small upper limit on the radial velocity of the disc at the
star-disc boundary appears to favour the viscous 
decretion disc model, but does not rule out other theories.

Whatever the mechanism which produces the disc, its properties must be
very similar to the disc model of Waters (1986); with the new limit on
the radial velocity (Porter 1998), the 
mass-loss rate in the disc is similar to that in the fast wind over
the polar regions (also see Okazaki 1997). It is noted that as long as
the viscosity is suitably defined then the disc model
may be described using identical mathematics as accretion discs
(e.g. Pringle 1981) -- indeed for the viscous
decretion discs (Lee \etal\ 1991) a viscosity is
introduced from the outset.
The main difference between the accretion discs and the decretion
discs here (viscous or not) is essentially the boundary conditions --
here the material drifts outwards; $v_r >0$ and angular momentum is
supplied {\em from} the star, possibly by pulsations (see Osaki 1986).

\subsection{Optical depth of the disc}

The model of Waters (1986) is used to represent disc structure.
It is assumed to have a constant opening angle $\theta$, and has
a density profile of $\rho = 10^{-11} \rho_0 (r/\rstar)^{-n}$g
cm$^{-3}$, where $\rho_0$
is the density at the inner edge measured in $10^{-11}$g cm$^{-3}$.
The disc temperature is assumed to be $T_{d} = 0.5T_{\rm eff}$.
It is also assumed to be isothermal, although this approximation
may be easily relaxed (see Waters 1986 for a discussion on
non-isothermality).

From this the optical depth of the disc may be calculated at a
frequency $\nu$ at which $\nu F_\nu$ at the disc temperature $T_d$ is
maximum (where $F_\nu$ is the Planck function). 
Using cylindrical geometry, the optical depth of the disc $\tau_\nu$
at a radius $R$ 
is calculated from the prescription in Waters (1986). Using his
notation 
\begin{equation}
\left.
\begin{array}{c}
\tau_\nu(R) = X_\lambda\ X_{\ast d}\ C\ R^{-2n + 1} \\
  \\
X_\lambda  = \lambda^2 
\frac{\left(\displaystyle {1 - {\rm e}^{-h\nu / kT_d }}
\right)}{\left(\displaystyle {\frac{h\nu}{kT_d}}\right)} \left( g_{ff} + g_{fb}
\right) \\
 \\
X_{\ast d} = 4.923\times 10^{13} {\displaystyle {\frac{\bar{z^2}}{T_d^{3/2}
{\mu^2}}}} \gamma
\rho_0^2 \left(\frac{\rstar}{\rsun}\right) \\
  \\
C = \displaystyle {\int}^{\theta}_0 {\rm cos}^{2n-2}y\ dy
\end{array}
\right\}
\end{equation}
where $\lambda$ is the wavelength in cm, $g_{ff}$ and $g_{fb}$ are the
Gaunt factors for free-free and free-bound emission respectively,
$\bar{z^2}$ is the mean squared atomic charge, $\gamma$ is the ratio
of the number of electrons to ions, and $\mu$ is the mean atomic mass.
The opening angle of the disc is $\theta$
and the rest of the symbols take their usual meanings.

Using a selection of model stars spanning the main-sequence B star
range various quantities for the model can be calculated (Table 1).
The wavelength $\lambda_m$ at which $\nu F_\nu$ is maximal is
displayed in column 7 of Table 1.
To calculate the optical depth, it has been assumed that the base density
in the discs for all the stars is identical: $\rho_0 = 1$.
There seems to be little
correlation between the spectral type of the star, and disc tracers
such as equivelent width of H$\alpha$ (see e.g. van Kerkwijk \etal\
1995), and so the assumption of a common density normalization for all
the stars is unlikely to be seriously in error.
The calculation has also assumed that $n = 2.5$, and $\mu = 1.6^{-1}$.

Although the disc is optically thick in line transitions out to several
tens of stellar radii
(e.g. Chokshi \& Cohen 1987, Kastner \& Mazzali 1989), eq.1
does not yield $\tau >1$ for $\lambda_m$. Therefore the wavelength at
which the optical depth is greater than unity $\lambda_{\tau =1}$ at
the inner edge of the disc is calculated, as is the fraction of
emission at that wavelength to the maximal emission $\nu F_\nu$. 
These quantities are presented in columns 8 and 9 of Table 1.
The fraction of the disc emission at $\lambda_{\tau = 1}$ is small for
the earliest B star, rising to 60\% for the latest spectral type
considered. 
The effect of this partial optical thickness of the disc is
likely to be on the magnitude of the warp, should the disc become
unstable (see below).
The discs around the earliest (latest) B stars will generate the 
smallest (largest) amplitude warp.
Clearly, if the disc is to become unstable to radiation-induced
warping, it must be unstable to these perturbations very close to the
star. 

%-----------------------------------------------------------------------%

\begin{table*}
\caption{Stellar parameters for the stars considered from
Schmidt-Kaler (1982).
The mass loss rates (in \msunyr) come from the relation by Garmany
\etal\ (1981).
$\lambda_m$ is the wavelength at which $\nu F_\nu$ at the temperature of
the disc ($T_d = 0.5T_{\rm eff}$) is maximal. The wavelength at which
the disc has an optical of unity at the inner edge is $\lambda_{\tau =
1}$. The fraction of disc emission at $\lambda_{\tau = 1}$ to that at
$\lambda_m$ is listed in the ninth column.
The tenth column gives
the growth timescale $t_\Gamma$ for the warping instability at $r =
\rstar$ from eq.7. The final column gives the bending (from eq. 12) at
which the wind-disc interaction becomes significant.} 
\begin{tabular}{ccccccccccc}
Spectral & \mstar  & \rstar  & log\lstar & $T_{\rm eff}$
&log$\mdot_w$ & $\lambda_m$ & $\lambda_{\tau = 1}$ &  emission
& $t_\Gamma$ & $\psi-\theta$ \\
type    & (\msun)  & (\rsun) &  & ($10^4$K)  &  &  ($\mu$m)  &
($\mu$m) & fraction & (days)
& (degrees) \\ \hline 
 B0     & 17.5 & 7.4 & 4.81 & 33.9 & -7.5  & 0.22 & 1.29 & 0.04 & 0.5 & 47 \\ 
 B1	& 13.0 & 6.4 & 4.31 & 27.5 & -8.4  & 0.27 & 1.23 & 0.08 & 0.9 & 40 \\ 
 B2	& 9.8  & 5.6 & 3.85 & 22.4 & -9.2  & 0.33 & 1.19 & 0.15 & 1.7 & 35 \\ 
 B3     & 7.6  & 4.8 & 3.43 & 19.0 & -9.8  & 0.39 & 1.18 & 0.22 & 2.6 & 30 \\ 
 B4	& 6.4  & 4.2 & 3.14 & 17.3 & -10.4 & 0.42 & 1.20 & 0.26 & 3.3 & 26 \\ 
 B5	& 5.5  & 3.8 & 2.89 & 15.7 & -10.8 & 0.47 & 1.21 & 0.31 & 4.3 & 24 \\ 
 B6	& 4.8  & 3.5 & 2.67 & 14.4 & -11.2 & 0.51 & 1.21 & 0.37 & 5.4 & 22 \\ 
 B7	& 4.2  & 3.2 & 2.45 & 13.3 & -11.6 & 0.55 & 1.22 & 0.43 & 6.7 & 20 \\ 
 B8     & 3.4  & 3.0 & 2.28 & 11.2 & -11.8 & 0.67 & 1.18 & 0.60 & 8.0 & 18
\end{tabular}
\end{table*}

%-----------------------------------------------------------------------%

\section{Disc warping}

The ciriterion that radiation-driven warping has been derived by 
Pringle (1996) and Malony, Bagelman \& Pringle (1996).
The typical
timescale for the radiation torque $t_\Gamma$ and the viscous
timescale for damping warps $t_{\nu_2}$ are:
\begin{equation}
t_{\nu_2} = \frac{2 R^2}{\nu_2}, \ \ \
t_\Gamma = 12\pi G^{1/2} c
\left( \frac{\Sigma \mstar^{1/2} R^{3/2}}{\lstar} \right)
\end{equation}
where $\nu_2$ is the $(R, z)$ component of the disc viscosity (see
e.g. Pringle 1992) \mstar and \lstar are the mass and luminosity of
the star, $G$ and $c$ are the gravitational constant and speed of
light respectively, and $\Sigma$ is the disc surface density.
For instability, Pringle (1996) estimated that $t_{\nu_2} > 2\pi
t_\Gamma$, a result which has been found consistent with numerical
simulations (Pringle 1997, Armitage \& Pringle 1997).

To make further progress on this problem for {\em decretion} type
discs, however, the angular momentum transport equation must be
examined again.
For a steady disc, a constant ${\cal C}$ is produced when the angular
momentum equation is integrated (Pringle's 1981 eq.3.8). For decretion
discs, this constant is the rate of angular momentum input
at the inner boundary of the disc. If the central star rotates with an
angular velocity of a
fraction $f_\ast$ of the Keplerian value, then the constant ${\cal C}$ should
become ${\cal C} = (1 - f_\ast) \mdot R^2 \Omega$.
This leads to a modification of Pringle's (1981) eq. 3.9:
\begin{equation}
\nu_1 \Sigma = \frac{\mdot}{3\pi} \left[ 1 - (1-f_\ast)
\left(\frac{R_\ast}{R} \right)^{1/2} \right],
\end{equation}
where $\nu_1$ is the $(R, \phi)$ component of viscosity.

With this and the instability criterion eq.2 then becomes
\begin{equation}
\left( \frac{R_\ast}{R}\right)^{1/2}
- (1 - f_\ast) \left( \frac{R_\ast}{R}\right) -
\frac{L R_\ast^{1/2}}{4\pi G^{1/2} c \eta M^{1/2} \mdot} <0.
\end{equation}
This is essentially different from the expression of Armitage \&
Pringle (1997) (their eq.3) as the full expression of our eq. 3 has
been used instead of the approximation in the case of large
$R/R_\ast$. As the disc is (partially) optically thick only in
the inner parts then the full expression is required.

Setting eq.4 equal to zero then defines the critical radius
$R_c$ separating regions of instability and stability.
With the replacement $r = R_c/\rstar$, this becomes
\begin{equation}
\left.
\begin{array}{c}
r^{1/2} - \left(1 - f_\ast\right) - gr = 0 \\
   \\
g = \frac{\displaystyle {3.7\times 10^{-2}}}
{\displaystyle {\eta \mdot_{-10}}}
\left( \frac{\lstar}{\lsun} \right)
\left( \frac{\mstar}{\msun} \right)^{-1/2}
\left( \frac{\rstar}{\rsun} \right)^{1/2}
\end{array}
\right\}
\end{equation}
where $\eta = \nu_2/\nu_1$, and the mass loss rate $\mdot_{-10}$ is
measured in $10^{-10}\msunyr$.

This equation yields Armitage \& Pringle's result (their eq.5) if the
$(1 - f_\ast)$ term is absent. Eq.5 is a quadratic equation for 
$r^{1/2}$, and so it may produce zero, one or two solutions
depending on the value of $g$. The solution is
\begin{equation}
r = \left( \frac{ 1 \pm \sqrt{1-4g(1-f_\ast)}}{2g} \right)^2.
\end{equation}
If only one solution is produced, then it is the inner boundary
beyond which warping instabilities may occur; if two solutions are
produced (both with $r\ge 1$), then the region between them is {\em
stable} to the warping instability.
If $4g(1~-~f_\ast)~>~1$ then no real solutions are obtained. This occurs
when eq.4 is obeyed for all values of $r$, i.e. the whole disc may be
unstable to warping -- in this case the warping occurs everywhere the disc
is optically thick.

To calculate the function $g$ the mass loss rate in the disc and
$\eta$ needs to be supplied. 
This is problematical for both quantities. 
As the viscosity may not be isotropic (see Pringle 1992), $\eta$ may
not necessarily be unity. 
Also, although the disc density may be
derived via observations of Be star disc continuum emission in the IR
(see e.g. Waters 1986) reasonably accurately, the radial velocity in
the disc, and hence the mass-loss rate is unknown. 
Decretion disc models produce low radial velocities (see e.g. Okazaki 1997),
and from angular momentum considerations the radial velocity at the
stellar surface must be $v_r \ltappeq 0.01$\kms (Porter 1998, but see
Porter's discussion).
Low velocities such as these imply mass-loss rates in Be star
discs similar to those for the radiation-driven wind over the poles
(Okazaki 1997). 
However, if the disc {\em is} a viscous decretion disc, then there is
no {\it a priori} reason that its mass loss rate and the fast wind's
mass loss rate should be the same as two different mechanisms are acting.

Clearly, there is some uncertainty here. 
To keep the calculation general, an ``ignorance'' parameter
$\Delta$ is introduced such that $\eta \mdot_{-10}$ is replaced by
$\Delta \eta \mdot_w$, where $\mdot_w$ is the mass-loss rate of the
fast wind over the pole (taken to be that from Garmany \etal's 1981 emperical
relation; column 6 Table 1).  
With this
introduction, the function $4g(1 - f_\ast)$ is calculated for the sample
of Be stars and is shown in fig.1.
\begin{figure}
\begin{picture}(100,270)
\put(0,0){\includegraphics{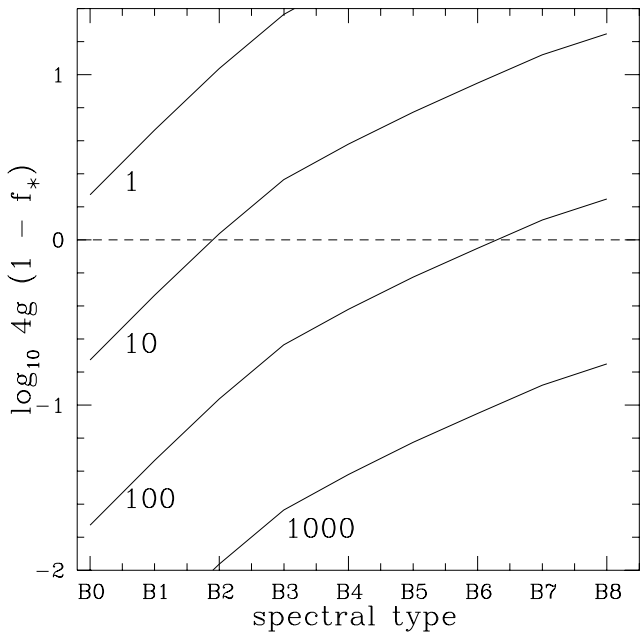}}
\end{picture}
\caption{The function $4g(1-f_\ast)$ defined in the text. The four
lines are labelled with the different values of $\Delta$ they
correspond to. The dotted line is the limit above which no real
solutions are obtained for eq.6.}
\end{figure}

Fig.1 shows that unless $\Delta$ is large, the disc is unstable for
all radii where it is optically thick. If the mass loss rate in the
disc is large, or the viscosity is very anisotropic, then eq.6 does
have two real solutions, and a region of the disc will be stable. 
It is found that the inner solution, where it exists, is located
within the star ($r < 1$), and
therefore only the second solution is meaningful. This outer solution
then provides a minimum radius at which the disc becomes unstable. 
As an example calculation, the outer solutions are calculated, where they
exist and $r > 1$ for values of $\Delta = 100, 1000$ and are displayed
in fig.2. 
\begin{figure}
\begin{picture}(100,270)
\put(0,0){\includegraphics{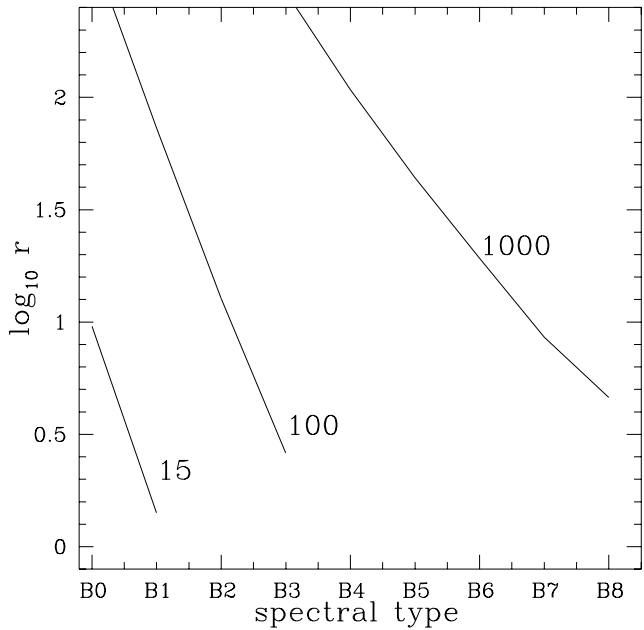}}
\end{picture}
\caption{The outer solution to eq.5, for the labelled values of
$\Delta$. There are no solutions which have $r > 1$ for $\Delta
\ltappeq 12$ for the stellar parameters in Table 1.}
\end{figure}

%-----------------------------------------------------------------------%
\subsection{Growth and precession rates}

How fast do the unstable modes grow when they are excited? The growth
timescale from eq.2 $t_\Gamma$ in years is
\begin{equation}
t_\Gamma = 1.4 
\left(\!\frac{\rstar}{\rsun}\!\right)^{\!\frac{5}{2}}\!\!
\left(\!\frac{\mstar}{\msun}\!\right)^{\!\frac{1}{2}}\!\!
\left(\!\frac{\lstar}{\lsun}\!\right)^{\!-1} \!\!\!
\rho_0\ {\rm tan}\theta\ 
\left(\!\frac{R}{\rstar}\!\right)^{\!\frac{5}{2} - n}\!\!.
\end{equation}
The precession timescale of the warp is
$T \approx 4\pi t_\Gamma$ which, when $R/\rstar \approx 1$, $\rho_0 \approx
1$ and tan$\theta \approx 0.1$ becomes
\begin{equation}
T \approx 1.7
\left(\!\frac{\rstar}{\rsun}\!\right)^{\!\frac{5}{2}}\!\!
\left(\!\frac{\mstar}{\msun}\!\right)^{\!\frac{1}{2}}\!\!
\left(\!\frac{\lstar}{\lsun}\!\right)^{\!-1} \ {\rm yr}.
\end{equation}
The growth timescales are displayed in column 10 of Table 1. 
It should
be pointed out that these timescales are short, typically ${\cal
O}$(days). Consequently, precesion timescales are ${\cal
O}$(months). The observational implications of these timescales
is returned to later in the discussion.

%----------------------------------------------------------------%

\section{Disc--fast wind interaction}

Once the disc starts to warp, it will interact with the fast,
radiatively driven wind from the star (which has radial velocities of $\sim
10^3$\kms, see e.g. Snow 1981). 
A bending mode will present a
non-negligible area to the fast wind and the fast wind will
interact via its ram-pressure.
In order to attempt to describe the evolution of the warp in the
optically-thin regions of the disc where the radiatively-driven
warping mechanism does not operate, some typical timescales are now
calculated.  

\subsection{Latitudinal timescale and magnitude of the warp}

The fast wind will interact with the warped disc and attempt to
deflect the warp toward the plane of neutral stability -- the
equatorial plane.
This interaction is complex, however, as it depends critically on
the amplitude, (or local inclination) of the warp, and is therefore a
function of the warp itself. 
The timescale for this deflection is denoted as the wind-disc timescale
$t_{wd}$ and is now estimated. 
The local disc opening angle is $\psi$,
(for bends $\psi$ is a monotonically increasing function of $R$), 
and the opening angle of a stationary disc is
$\theta$ (see Fig.1, Porter (1997). 
The velocity of the wind perpendicular to the disc
surface is $v_\perp = v_w {\rm sin}(\psi - \theta)$, where $v_w$ is
the velocity of the fast wind (which is assumed not to have a meridional
component -- although see Bjorkamn \& Cassinelli 1993, and Owocki, Cramner
\& Gayley 1996)
The ram pressure directed toward the equatorial plane is then $\rho_w
v_\perp^2$ where $\rho_w$ is the density of the wind.
This pressure acts on an area of $\Phi R\ \delta\!R / {\rm
cos}\psi$, where $\Phi$ is the angle over which the warp exists (e.g. for a
simple inclined disc, $\Phi = \pi$).

The acceleration $g_{wd}$ toward the equatorial plane provided by this
interaction is equal to the force due to the ram-pressure (resolved into the
$z$ direction),
divided by the mass $dm$ on which it acts
$dm = \Phi R\ \delta \!R \rho_d H$, where $\rho_d$ is the disc density
and $H = R\ {\rm tan}\theta$ is the scale height of the disc. Hence,
the acceleration is:
\begin{equation}
g_{wd} = \left( \frac{\rho_w v_w^2}{\rho_d} \right)
\frac{{\rm sin}^2(\psi - \theta)}{R {\rm tan}\theta}.
\end{equation}

The timescale for ``flattening out'' warps by this wind-disc
interaction depends on how far the disc has moved from the equatorial
plane $\delta H = R({\rm tan}\psi - {\rm tan}\theta)$. Assuming that
both $\theta$ and $\psi$ are small, then the timescale is $t_{wd}^2
\sim R (\psi - \theta)/g_{wd}$, or
\begin{equation}
t_{wd} = R
\left( \frac{\rho_d}{\rho_w v_w^2}\right)^{\frac{1}{2}}
\left( \frac{\theta}{\psi - \theta} \right)^{\frac{1}{2}}.
\end{equation}
Setting $\rho_w = \mdot_w / 4\pi R^2 v_w$, and again using $\rho_d =
10^{-11}\rho_0 (R/\rstar)^{-n}$, this timescale (in seconds) becomes
\begin{equation}
t_{wd} = 6.8\times 10^4 \!
\left(\!\frac{R}{\rstar}\!\right)^{\!2-\frac{n}{2}}\!\!\!
\left(\!\frac{\rstar}{\rsun}\!\right)^2\!\!
\frac{\rho_0^{\frac{1}{2}}}{v_8^{\frac{1}{2}} \mdot_{w}^{\frac{1}{2}}}\!
\left(\!\frac{\theta}{\psi - \theta}\!\right)^{\frac{1}{2}}
\end{equation}
where $v_8$ is the wind velocity in $10^8$cm s$^{-1} (10^3$\kms) and
the wind mass-loss rate $\mdot_w$ is measured in $10^{-10}$\msunyr.

This expression may now be used to estimate the magnitude of bending
modes which may grow via the effect of radiation before they are
seriously peturbed by the wind-disc interaction. This is done by
calculating the bending ($\psi - \theta$) at which the wind-disc
timescale $t_{wd}$ is equal to the radiatively-driven warping timescale
$t_\Gamma$. 
This yields
\begin{eqnarray}
\theta(\psi - \theta) = 1.6\times 10^{-6}\!\! & & \!\!\! 
\left(\!\frac{\mstar}{\msun}\!\right)^{-1}\!\!
\left(\!\frac{\rstar}{\rsun}\!\right)^{-1}\!\!
\left(\!\frac{\lstar}{\lsun}\!\right)^{2} \times \nonumber \\
 & & 
\frac{1}{\rho_0 v_8 \mdot_{w}}
\left(\!\frac{R}{\rstar}\!\right)^{n-1}.
\end{eqnarray}
Assuming $v_8 \approx 1$, and values of $\mdot_w$ from Garmany \etal's
(1981) relation, 
the values of $\psi - \theta$ are listed in column 9 of Table 1 at
the inner edges of the disc ($R/\rstar \approx 1$), for $\theta = 0.1$rad --
it is clear that large warps can develop before they are seriously
affected by the wind-disc interaction. In fact for the early B stars
the approximation that $\psi$ is small is severely strained.

\subsection{Is the warp driven to larger radii}
Clearly, the fast wind will attempt to force the warp to larger
radii. The effectiveness of this depends on the ratio of the radial
momentum flux in the wind to the gravitational attraction of the
disc. If this ratio is greater (less) than unity then the warp will
(will not) be driven outwards.

The wind's momentum flux is simply the product of mass-loss rate and
the wind velocity, weighted by the ratio of the solid angle of interaction to
$4\pi$steradians. The area of interaction is $\Phi R \delta R {\rm
sin}(\phi-\theta)/ ({\rm cos}\theta\ {\rm cos}\psi)$, 
and so the wind's momentum flux ${\cal F}$ is
\begin{equation}
{\cal F}_m = \mdot_w v_w \left( \frac{\delta R {\rm
sin}(\psi-\theta)}{ 2R\ {\rm cos}\theta\ {\rm cos}\psi} \right).
\end{equation}
The gravitational interaction of the star and the mass over which the
wind's momentum flux acts is
\begin{equation}
{\cal F}_g = \frac{G \mstar}{R^2}dm = \frac{\Phi G \mstar \rho_d H
\delta R}{R}.
\end{equation}

Scaling the wind's mass-loss rate and velocity to $10^{-10}\msunyr$,
and $10^8$cm s$^{-1}$ respectively, and using the power law expression
for the disc, the ratio of the two terms above is
\begin{eqnarray}
\frac{{\cal F}_m}{{\cal F}_g} = 5.4\times 10^{-4} \mdot_w v_8 
\left(\!\frac{\mstar}{\msun}\!\right)^{-1}\!\!\!\! & & \!\!\!
\left(\!\frac{\rstar}{\rsun}\!\right)^{-1}\!\!
\left(\!\frac{R}{\rstar}\!\right)^n \times \nonumber \\
 & &
\left(\frac{{\rm sin}(\psi-\theta)}{{\rm sin}\theta\ {\rm cos}\psi} \right).
\end{eqnarray}

In the last section it was found that $\psi$ will be tens of degrees
when the wind-disc interaction takes placeand so the trigonometric
term will numerically be $\sim$10s. Therefore, it is clear that the
fast wind {\em cannot} drive the warps to large radii. 

%-----------------------------------------------------------------------%

\section{Discussion}

From the preceding sections it seems that Be star discs {\em do} posess
the correct attributes to enable them to become unstable to
radiatively-driven warping.
If this is so, then they should be observable. 
In fact it may well be that 
this is indeed the
case -- Hummel (1998) has indeed interpreted line-profile variations in 
$\gamma$Cas 
(B0IVe) and 59~Cyg (B1.5Vnne) to be due to a warped inner disc. This
adds a large degree of credibility to the above approach.
Hummel {\em did} consider the radiatively-induced warping mechanism
as an explanation for the observations, but rejected the
mechanism as he found a prohibitively large mimimum radius for the
instability to operate ($10^7\rstar$). The inconsistency between his
result and that 
presented here lies in the application of accretion disc theory
to {\em decretion} discs of Be stars. 
(This involved approximating the accretion efficiency $\epsilon =
L/\mdot c^2$ by
the ratio of the Schwarzschild radius to the stellar radius, leading
to an error of $\sim 5\times 10^3$. This value is squared to yield the
inner radius for instability, which gives an overestimate of $\sim
2\times 10^7$, bringing Hummel's value into line with that here.)

A feature of the development of the warps which is specific to Be
stars is the action of the central
star's radiatively driven wind. This introduces an extra (warp
dependant) effective viscosity leading to a different evolution of
bending modes.
 
\subsection{Development of a warp}
In the region of the disc where it is optically thick, radiatively
induced warps grow on the timescale $t_\Gamma$. These warps are
unmolested until they become large amplitude bends (defined from
eq.12). At this point the wind-disc interaction will start to 
domiate over the induced warping, 
and the bending mode will oscillate about the
equatorial plane under the action of the fast wind's ram pressure. 
If it is still within the optically thick part of the
disc, then the combined effects of the induced warping, and the
wind-disc interaction will limit the oscillation's amplitude. However,
when the 
mode propagates out of the optically thick part of the disc, then the
radiation-induced warping becomes inoperable, and the evolution of
the mode is determined entirely by the wind-disc interaction.
It has been found that the warp will not be driven to large radii
under this interaction, and will only propagate outwards via the
action of viscosity (for decretion discs) or advection.
However, it is difficult to determine whether the warp will be
amplified or whether it will decay due to the effects of the wind --
further work on this point is necessary (Porter, in  preparation).

\subsection{Observational signatures}
The timescales for the growth of radiatively-driven warps are shown in
Table 1. Although modes may be excited over this short timescale, it
is not necessarily the timescale over which observed disc tracers
indicate the warp. 
Once a
mode propagates outward beyond the optically thick part of the disc,
then the timescale for its evolution and precession will be determined
by the wind-disc timescale (eq.11).
Consider a bending mode grows to its maximum amplitude where the
wind-disc interaction becomes effective. Once it has propagated out in
the disc and its amplitude has decreases, then the timescale the mode
evolves at falls (see eq.11). 

The observational properties of such a mode are not defined simply by
the optically thick part of the disc defined in \S2. Typically, the
hydrogen lines form within several tens of stellar radii 
(e.g. Hanuschik \etal\ 1988)
and it is these which are commonly used to infer properties about the disc.
Consequently, the warps inferred from the observations are likely
to have longer timescales than the growth timescales,
and precession rates in eq.7 and 8. 

For example, consider a warp which had propagated to 10\rstar, and has
$\psi - \theta = 0.1$. The precession timescale for the wind-disc
interaction $T = 4\pi t_{wd}$, ranges from 170 days for a B0 star to
4200 days for a B8 star, considerably longer than the timescale over
which the mode grows. Therefore it is likely that observationally
derived precession timescales will range from several hundreds, to
several thousand days.
This is broadly consistent with Hummel (1998), who ascribes timescales of
$\sim 1000$days for precession of the discs around $\gamma$~Cas and 59~Cyg.

%-----------------------------------------------------------------------%
\section{Conclusion}

The theory of radiatively-induced warping of discs has been applied to
Be stars and found to be important, especially for late B stars. 
The timescales associated with the growth of warps is short ${\cal
O}$(days), although these may not be the relevant timescales over
which the modes are examined observationally.
It is noted that warping has been observed in Be star discs already
(Hummel 1998), and that the observationally derived precession
timescales fall in the range in which the simple argument predicts.

A scenario has been presented in which radiatively-driven warps are
created in the central parts of the disc and propagate outwards under
the influence of the interaction of the disc and fast stellar wind.

%-----------------------------------------------------------------------%
\section*{Acknowledgements}
I thank Dr. J. Pringle for reading and suggesting improvements on the
first draft of this paper.

JMP is supported by a PPARC postdoctoral research assistantship.

%-----------------------------------------------------------------------%
{}

\end{document}